# Measuring Creativity in the Age of Generative AI: Distinguishing Human and AI-Generated Creative Performance in Hiring and Talent Systems

Yigal Rosen[1], Ph.D.

*Massachusetts Institute of Technology & Ignis AI*

and

Ilia Rushkin, Ph.D.

*Ignis AI*

**Abstract**

Generative AI is rapidly transforming how organizations create value and evaluate talent. While large language models enhance baseline output quality, they simultaneously introduce ambiguity in assessing human creativity, as observable artifacts may be partially or fully AI-generated. This paper reconceptualizes creativity as a distributional and process-based property that emerges under shared constraints and competitive incentives. We introduce a quantitative framework for measuring creativity as novelty in synthesis, operationalized through idea generation and idea transformation within embedding space. Empirical evaluation demonstrates that the proposed metrics align with intuitive judgments of creativity while capturing distinctions that surface-level quality assessments miss. We further identify a structural shift toward bimodal distributions of creative output in AI-mediated environments, with implications for hiring, leadership, and competitive strategy. The findings suggest that in the age of generative AI, distinctiveness rather than fluency becomes the primary signal of human creative capability.

## 1. Introduction

Generative artificial intelligence has fundamentally altered the landscape of creative work. In domains ranging from education to enterprise decision-making, individuals increasingly rely on large language models (LLMs) to generate written content, analyze complex scenarios, and propose strategic solutions. While these systems enhance productivity and elevate baseline output quality, they simultaneously disrupt long-standing assumptions about how human capability is observed and evaluated.

---

[1] Correspondence concerning this research paper should be addressed to Dr. Yigal Rosen: yigal@ignisai.ai / yigal@mit.edu

In hiring and workforce development contexts, organizations traditionally infer creative competence from observable outputs such as essays, case analyses, or project deliverables. These artifacts have historically served as proxies for underlying cognitive ability. However, as generative AI becomes ubiquitous, this assumption is no longer tenable. Outputs are now frequently co-produced by humans and machines, and the relationship between observable performance and individual capability becomes increasingly opaque.

This shift creates a fundamental measurement problem. Organizations must now distinguish between outputs that reflect genuine human creativity and those that primarily reflect the statistical priors of generative models. Failure to do so risks conflating AI-enabled fluency with human distinctiveness, thereby undermining talent identification and strategic decision-making (Doshi & Hauser, 2024; Kleinberg & Raghavan, 2021).

## 2. Creativity as a System-Level Phenomenon

Traditional accounts of creativity often focus on individual cognitive processes, such as divergent thinking or associative recombination (Guilford, 1967; Mednick, 1962). While these perspectives remain valuable, they are insufficient to explain the dynamics of creativity in contemporary socio-technical systems. In particular, they do not account for the ways in which tools, knowledge systems, and participation structures shape the realization of creative potential.

To address this limitation, we conceptualize creativity as a multiplicative function of five interdependent factors: idea generation, idea transformation, knowledge base, tools, and participation. This formulation captures the intuition that creativity scales not only through individual cognition but also through the expansion of combinatorial possibilities enabled by technological and institutional infrastructures.

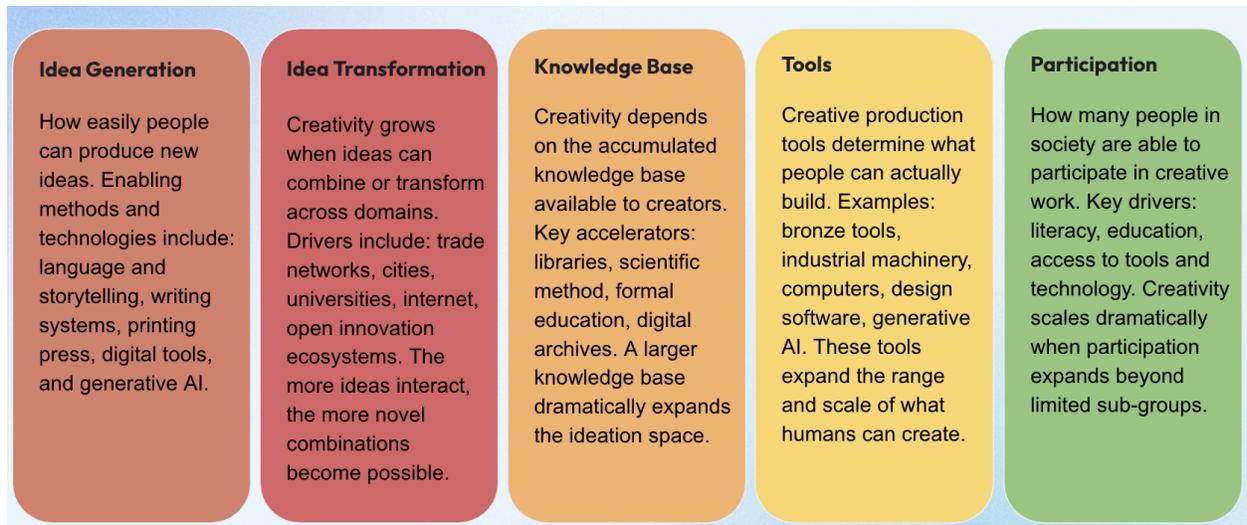

Figure 1. Creativity as a multiplicative function of five interdependent factors

Historical evidence supports this view. Major inflection points in human creativity—such as the invention of writing, the rise of scientific institutions, and the emergence of digital networks—did not fundamentally alter human cognitive architecture. Rather, they amplified the range and scale of what humans could imagine, recombine, and produce (Mokyr, 2002; Arthur, 2009). This aligns with theories of recombinant growth, which emphasize the combinatorial expansion of ideas as a driver of innovation (Weitzman, 1998).

Generative AI represents a new phase in this trajectory. By dramatically increasing the speed and breadth of idea generation, these systems expand the accessible search space of possible solutions. At the same time, however, they introduce new forms of convergence, as outputs cluster around high-probability regions of the model's learned distribution (Holtzman et al., 2020; Li et al., 2024). This dual effect—expansion of possibility coupled with compression of diversity—lies at the heart of the contemporary creativity paradox.

### 3. The Collapse of Traditional Evaluation Signals

The widespread adoption of generative AI has destabilized traditional signals used to evaluate human capability. In many contexts, applicants and employees now use LLMs to produce artifacts that are indistinguishable, in surface quality, from those produced by highly skilled individuals. As a result, evaluators face increasing difficulty in attributing observed performance to underlying human competence.

This challenge is compounded by the tendency of generative systems to produce outputs that are statistically typical rather than truly novel. Research on text generation has documented phenomena such as mode collapse and degeneration, whereby outputs concentrate around common patterns even when prompted for diversity (Holtzman et al., 2020; Srivastava et al., 2023; Zhang et al, 2024). Similarly, recent studies have shown that generative AI can enhance individual creativity while simultaneously reducing collective diversity, increasing the risk of algorithmic monoculture (Doshi & Hauser, 2024; Wu et al., 2025).

In this environment, traditional evaluation methods—such as rubric-based scoring or subjective judgment of output quality—become increasingly unreliable. They are prone to overvaluing fluency and coherence while underdetecting convergence and lack of originality. Consequently, organizations risk selecting candidates who are proficient in leveraging AI tools but lack the capacity for differentiated thinking.

## 4. A Distributional View of Creativity

To address these limitations, we propose a distributional view of creativity. Rather than evaluating outputs in isolation, we assess them relative to a population of competing responses generated under similar conditions. In this framework, creativity is defined as meaningful divergence from the distribution of available solutions.

This perspective aligns with recent work in economics and computational social science, which emphasizes the role of competition and incentive structures in shaping diversity outcomes (Raghavan, 2025). It also resonates with foundational theories of organizational learning, particularly the balance between exploration and exploitation in the search for novel solutions (March, 1991).

A key implication of this view is that creativity is inherently contextual. The same response may be considered creative in one distribution but conventional in another. Consequently, measurement must account for the structure of the solution space and the relative positioning of individual contributions within it.

## 5. Bimodal Creativity in AI-Mediated Environments

Building on this distributional framework, we identify a structural shift in the distribution of creative outputs under conditions of shared access to generative AI. Specifically, we observe the emergence of a bimodal distribution, characterized by two distinct clusters.

The first cluster consists of outputs that align closely with the default patterns of the generative model. These outputs are typically high in fluency and coherence but exhibit limited originality. The second cluster consists of outputs that diverge meaningfully from these patterns, reflecting human-driven recombination, reframing, or synthesis. Between these clusters lies a relatively sparse region, indicating a lack of intermediate solutions.

This bimodal structure is consistent with prior observations of distributional collapse and reduced diversity in generative systems (Li et al., 2024; Wu et al., 2025). It also reflects broader dynamics observed in innovation systems, where shared tools and incentives can lead to convergence unless deliberate mechanisms are introduced to promote differentiation (Kleinberg & Raghavan, 2021).

For organizations, this implies a fundamental shift in the evaluation of talent. The objective is no longer to identify individuals who can produce high-quality outputs, but rather those who can produce outputs that are meaningfully different from what the model would generate by default.

## 6. Quantifying Creativity as Novelty and Entropy in Synthesis

To operationalize this concept, we define creativity as novelty in synthesis, modeled as the product of idea generation and idea transformation. Given a set of premise statements (abstract ideas, disparate facts, concepts from different domains, etc.), an inference statement is produced by the test subject in response. Idea generation captures the extent to which a response introduces novel elements, while idea transformation captures the degree to which these elements are integrated across premises.

This formulation builds on established theories of creativity, including divergent thinking as the expansion of the idea space (Runco & Acar, 2012) and associative recombination as the basis of creative insight (Mednick, 1962). It also aligns with research on ambidextrous leadership, which emphasizes the interplay between exploratory and exploitative processes in innovation (Rosing et al., 2011).

We implement this framework using a geometric representation in embedding space, where premises and responses, as well as their subelements, are represented as high-dimensional vectors. Considering the projection of a response (or a response subelement) onto a cone spanned by the premises (or their subelements), novelty is quantified by the norm of the complement to the projection, while transformation is quantified using entropy-based measures that capture the breadth and balance of contributions across premises. The two measures—novelty and entropy—are then combined into a single creativity score by means of a mathematical model that allows tunable meta-parameters.

Importantly, this approach does not rely on opaque generative processes or subjective scoring. Instead, it provides a transparent and consistent numerical measure grounded in first principles. While it does not claim to capture all aspects of creativity, it offers a valid and operationalizable definition that can be applied across diverse contexts.

## 7. Empirical Evaluation

We evaluate the proposed framework using a synthetic AI-generated dataset of activities (sets of premises) and responses of varying levels of creativity. This provides us with a labeled set of activities and responses. Generation is done with a general-purpose generative model with straightforward prompts, so that it serves as an approximation to human judgments of what is more or less creative. Thus, while we compare our model scores to the labels, the labels cannot be regarded as absolute truth, only an approximation to human-perceived creativity. Hence we are evaluating not only literal closeness (mean absolute error), but also correlation: Pearson rho and Kendall tau (a measure of ordinal agreement).

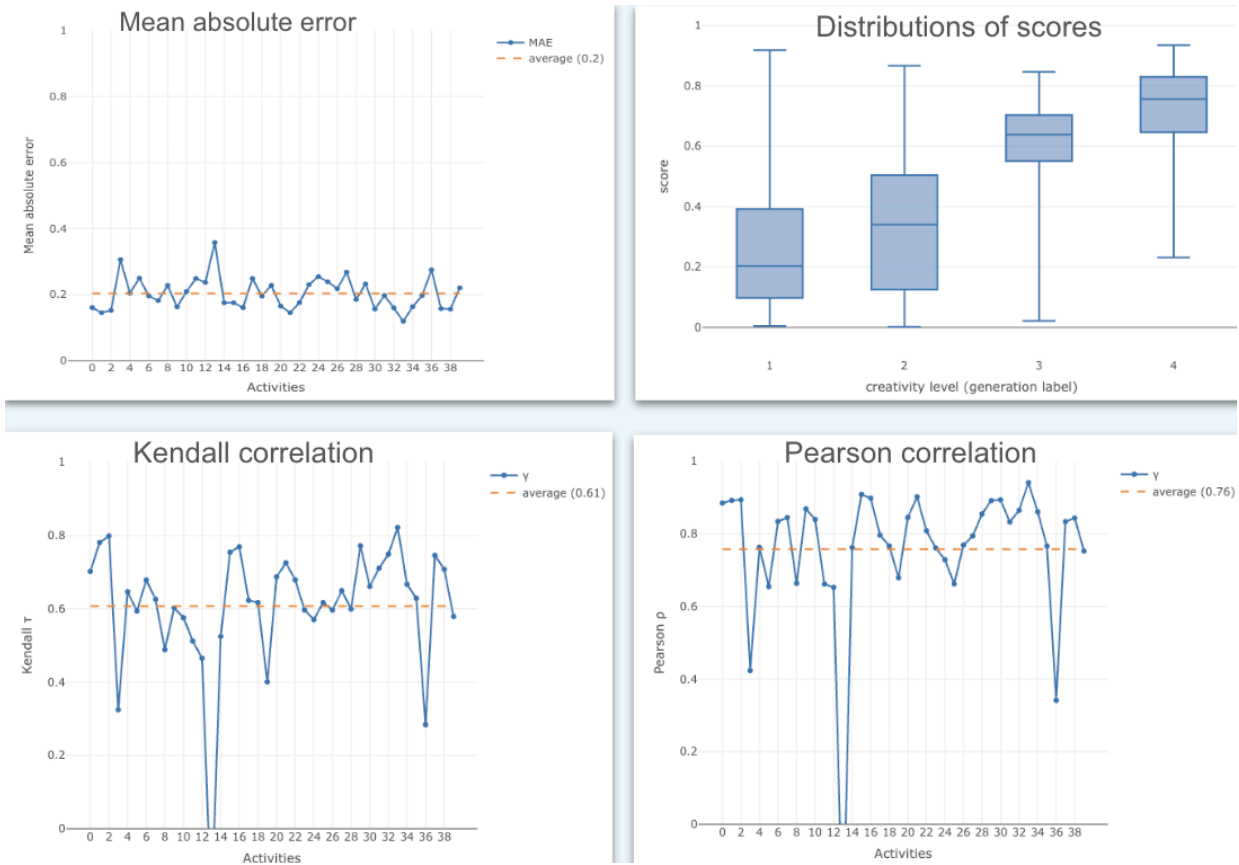

Figure 2. Preliminary results demonstrate that creativity can be measured reliably across assessment activities.

Mean absolute errors of 40 activities are typically low (with both the model outputs and the labels being on 0-to-1 scale): 0.20 average across activities. Kendall and Pearson correlation coefficients are typically high (0.61 and 0.76 averages across activities).

Score distributions are provided to illustrate the relationship: on the x-axis are the labels (converted to integers 0-4). The box-and-whiskers plots show the sets of model outputs for those responses. We see that they go steadily up with X and cover a large portion of the theoretically defined 0-1 range on the y-axis.

Results indicate low mean absolute error and substantial correlation between model scores and intuitive labels, suggesting that the framework aligns with common notions of perceived creativity. Moreover, the variability of metrics across activities provides us with further insights and serves as the beginning of an iterative convergent process: analyzing differences between activities with better and worse agreement metrics (like #33 vs. #13), we will iterate on the activity-generation process in order to further stabilize it and reduce the scatter. Tests on human-labeled data will also be done.

## 8. Implications for Organizations and Strategy

The findings of this study have significant implications for organizations operating in AI-mediated environments. First, they suggest that creativity signals have fundamentally shifted from output quality to distinctiveness relative to an AI baseline. Second, they highlight the need to distinguish between AI fluency and human creative competence, as the former does not guarantee the latter.

More broadly, the results indicate that competitive advantage will increasingly depend on the ability to identify and cultivate individuals who can operate outside the dominant modes of generative systems. This perspective aligns with broader economic analyses of technological change, which emphasize the importance of complementary human capabilities in the age of intelligent machines (Brynjolfsson & McAfee, 2014; Acemoglu & Johnson, 2023).

## 9. Conclusion

As generative AI becomes an integral part of creative production, the challenge of measuring human capability becomes both more complex and more critical. This paper proposes a framework for addressing this challenge by reconceptualizing creativity as a distributional property and introducing a quantitative method for its measurement.

The central insight is that in environments where everyone has access to the same generative tools, creativity is defined not by fluency but by distinctiveness. Organizations that fail to recognize this shift risk converging toward homogeneity and losing their capacity for innovation. Conversely, those that develop the ability to measure and cultivate differentiated thinking will be better positioned to thrive in the age of intelligent systems.